\documentclass[aip]{revtex4-1}

\begin{document}


\title{Pure radiation in space-time models that admit integration of the eikonal equation by the  separation of variables  method} 

\author{Osetrin Evgeny}
\email[]{evgeny.osetrin@gmail.com}
\affiliation{Tomsk State Pedagogical University\\ Russia, 634061, Tomsk, Kievskaya str.~60}

\author{Osetrin Konstantin}
\email[]{osetrin@tspu.edu.ru}
\altaffiliation[Also at]{Tomsk State Pedagogical University}
\affiliation{Tomsk State University\\ Russia, 634050, Tomsk,  Lenin av.~36}



\date{26 August, 2017}

\begin{abstract}
We consider space-time models with pure radiation, which admit integration of the eikonal equation by the method of separation of variables. For all types of these models, the equations of the energy-momentum conservation law were integrated. The resulting form of metric, energy density  and wave vectors of radiation as functions of metric for all types of spaces under consideration is presented. The solutions obtained can be used for any metric theories of gravitation.
\end{abstract}



\maketitle 

\section{Introduction}

At present, activity has increased in the field of studying realistic cosmological models, which is due to observational data on the accelerated expansion of the universe, the presence of isotropy disturbances in background cosmological radiation, and so on. If Einstein's theory of gravity is correct, then there must exist "dark" energy and "dark" matter, which interact gravitationally, but do not interact electromagnetically. The study of the properties of these exotic substances is based on astrophysical observations and theoretical modeling 
sometimes leads to exotic equations of state.

There is another option implemented, in which GR is an approximation to a more realistic theory of gravity, the search for which is also actively pursued. In any case, the situation needs new ideas, methods and tools for studying problems of cosmology and astrophysics.

In this research, we propose an approach to the modeling of space-time with pure radiation based on a combined approach with the following requirements:
\begin {itemize}
\item realistic theory of gravity is a metric theory, i.e. Gravity is modeled by a metric tensor, and free bodies and radiation move along the 
geodesic lines of space-time;
\item the law of conservation of energy-momentum of matter is satisfied;
\item to construct integrable models, we will use spaces that allow the integration of the eikonal equation by the method of separation of 
variables.
\end{itemize}
As will be shown later, for the space-time models that meet these requirements, one can integrate the equations of the energy-momentum 
conservation law and obtain expressions for the energy density and the wave vector of pure radiation  in the form of functional expressions in terms of functions included in the space-time metric. 
The explicit form of the field equations of any gravitation theory  is not required.

Solutions are written in coordinate systems that allow separation of variables in the eikonal equation. 
At the same time, part of the solutions 
obtained includes the use of an isotropic (wave-like) variable, which is used in modeling wave processes similar to electromagnetic or gravitational radiation (massless fields).

In this paper, we find and enumerate all types of models considered.  The results obtained can be used when comparing similar models in Einstein's theory of gravity and in various modified theories of gravitation.

When considering metric theories of gravity, then the motion of the test particles and radiation along geodesic lines, just as like the law of conservation of energy-momentum of matter, remain unchanged theory fundamentals:
\begin{equation} 
 \nabla^iT_{ij}=0,
\qquad
i,j,k=0...3,
\label{law_of_conservation}
\end{equation} 
where $T_{ij}$ is the energy-momentum tensor of matter.

Therefore, to study and compare modified theories of gravity, the use of space-time models that allow the existence of coordinate systems, where the eikonal equation for massless test particles (null geodesic line for the metric $g^{ij}$)
\begin{equation} 
g^{ij}S_{,i}S_{,j}=0,
\label{eq01}
\end{equation} 
or the Hamilton-Jacobi equation for test uncharged mass test particles (an analogue of the geodesic equations)
\begin{equation} 
g^{ij}S_{,i}S_{,j}=m^2,
\label{eq02}
\end{equation} 
or the Hamilton-Jacobi equation for a charged test particle in an electromagnetic field (with a potential $A_i$)
\begin{equation} 
g^{ij}(S_{,i}+A_i)(S_{,j}+A_j)=m^2,
\label{eq03}
\end{equation} 
allowing  integration by the method of complete separation of variables is of great interest (S is the action of the test particle in the Hamilton-Jacobi formalism, $S_{,i}=\partial S/\partial x^i$).

For these models, as it turns out, it is possible to integrate the differential equations of energy-momentum conservation law  (\ref{law_of_conservation}) for dust matter 
with the energy-momentum tensor of the form
\begin{eqnarray} 
&T_{ij}=\rho\, u_iu_j, &
\label{eq1}
\\
& u_iu^i=1,&
\label{eq_norm1}
\end{eqnarray} 
where $\rho$ -- energy density of dust matter and $u_i$  -- field of the four-velocity of dust matter.

The same situation and for pure radiation (in the high-frequency approximation of geometric optics)
with the energy-momentum tensor of the form
\begin{eqnarray} 
&  T_{ij}=\varepsilon\, {L}_i{L}_j, &
\label{eq1_1}
\\
&g^{ij} {L}_i{L}_i=0,&
\label{eq_norm0}
\end{eqnarray}
where $\varepsilon$ -- energy density and $L_i$  -- wave vector of radiation.

As a result, for the energy density of dust matter $\rho$ and radiation $\varepsilon$ , for the field of four-velocity of matter $u_i$ and for the wave vector of  radiation $L_i$ functional expressions can be obtained only through the space-time metric.

Thus, in these models, the use of field equations of specific theories of gravity can be reduced only to equations for the metric of space-time.

The spaces in which integration of the eikonal equation  (\ref{eq01}) is possible by the method of complete separation of variables are called {\bf conformally Stackel spaces~(CSS)}, in contrast to the {\bf Stackel spaces}, which admit complete separation of variables in the Hamilton-Jacobi equation for test uncharged masses  (\ref{eq02}).
In contrast to the case of the Hamilton-Jacobi equation for test mass particles, in the case of integration of the eikonal equation, the space-time metric admits an arbitrary conformal factor.

The task of this paper is to obtain a classification of functional expressions for the energy density of radiation and the wave vector by integrating the differential energy-momentum conservation law (\ref{law_of_conservation}) for pure radiation for all possible space-time models, where the eikonal equation  (\ref{eq01}) admits integration by the method of separation of variables.
An analogous problem for the case of Stackel spaces and dust matter was considered earlier (see ref.~\onlinecite{00}).

The theory of Stackel spaces for the case of the Lorentz signature was considered in the works of various authors (see, for example, refs.~\onlinecite{0}-\onlinecite{1}) and was formulated in the final form in the works of V.N. Shapovalov (see  refs.~\onlinecite{2}-\onlinecite{3}). We are based on the formulations of the theory and notation in (ref.~\onlinecite{4}).
The covariant condition of Stackel space is the presence of the so-called ''complete set'' of a commuting Killing vector and tensor fields, which satisfy some additional 
algebraic relations (see details in ref.~\onlinecite{4}).
The Stackel space metric tensor in privileged coordinate systems (where separation of variables is allowed) is determined by a 
set of arbitrary functions where each function depends only on one variable. 

{\bf Stackel space-times (SST)} application in gravitation theories (refs.~\onlinecite{5}--\onlinecite{13}) is based on the fact that exact integrable models can be developed for these spaces. The majority of 
well-known exact solutions is classified as SST (Schwarzschild, Kerr, Friedman-Robertson-Walker, NUT, etc.). It is important to note that 
the other single-particle equations of motion - Klein-Gordon-Fock, Dirac, Weyl -  admit a separation of variables  in SST. The same method can be 
used to obtain solutions in the  theories of modified gravity (refs.~\onlinecite{14}--\onlinecite{16}).






\section{Pure radiation in conformally Stackel space-times}

%
%

The Stackel spaces type is defined by a set of two numbers  $(N.N_0)$, where $N$ -- the number of commuting Killing vectors 
$Y^i_{(p)}$ ($p=1,N$) 
admitted by the ''complete set'' (the dimension of the Abelian group of space-time motions), and $N_0=N-rank|Y^i_{(p)}g_{ij}Y^j_{(q)}|$ -- is the number of null ("wave-like") variables in privileged coordinate 
systems. For 4-dimensional Stackel spaces of Lorentz signature $N=0...3$, $N_0=0,1$.

Stackel spaces metrics are conveniently written in privileged coordinate systems in a contravariant form.
The coordinates of space-time will be numbered from 0 to 3 by indices $i$, $ j$, $k$.
Ignored variables (on which the metric does not depend) will be numbered in indices $ p$, $q$, $r$ and non-ignored by Greek letters (indices) $ \mu $, $ \nu $, $ \sigma $.

The following notation will be used in the paper:
\begin{equation} 
P=\ln\left |\frac{{\varepsilon}^2}{\Delta^2 \det g^{ij}}\right |,
\end{equation} 
where $g^{ij}$ is the space-time metric, $\varepsilon$ is the energy density of radiation and $\Delta $ is the conformal  factor of metric $g_{ij}$.

For  conformally Stackel spaces the conformal factor of metric is an arbitrary function of all variables.

In the privlieged coordinate system, the wave vector of the radiation has a "separated" \, form, i.e. the corresponding covariant components of the wave vector depend only on one variable ${L}_i={L}_i(x^i)$:
\begin{eqnarray}
& L_0=L_0(x^0),
\quad
 L_1=L_1(x^1),
&\\
&
 L_2=L_2(x^2),
\quad 
L_3=L_3(x^3).
&
\nonumber
\end{eqnarray}
Later in the paper we obtain a functional form of the energy density and the wave vector of radiation  for all types of conformally Stackel space-times in privileged coordinate systems (where the separation of variables in the eikonal equation  (\ref{eq01}) is allowed).


\subsection{Conformally Stackel space-times (3.0) type}
Conformally Stackel space-times (3.0) type admit three commuting Killing vectors. In the privileged coordinate system, the metric of a conformally Stackel space of type (3.0) can be written in the following form:
\begin{equation} 
g^{ij}=\frac 1\Delta\pmatrix{1&0&0&0\cr 0&a_0&b_0&c_0\cr 0&b_0&d_0&e_0\cr 0&c_0&e_0&f_0},
\label{metric-3-0}
\end{equation} 
where  $\Delta=\Delta(x^0,x^1,x^2,x^3)$ and  $ a_0, b_0, c_0, d_0, e_0, f_0$  are functions of a variable $x^0$.

The wave vector of the radiation $L_i$ in the privileged coordinate system has the following "separated"\, form:
\begin{eqnarray}
&
{L}_0={L}_0(x^0),\quad {L}_1=\alpha,\quad {L}_2=\beta,\quad {L}_3=\gamma, 
&\\
&
\alpha, \beta, \gamma  -  const.
&\nonumber
\end{eqnarray}
Wave vector norm condition (\ref{eq_norm0}) takes the form:
\begin{equation} 
 \alpha^2a_0+2\alpha\beta b_0+\beta^2d_0+2\alpha\gamma c_0+2\beta\gamma e_0+\gamma^2 f_0=-{L}_0{}^2.
 \label{norm_condition_3-0}
\end{equation} 
From the equations of the energy-momentum conservation law (\ref{law_of_conservation}) we obtain:
\begin{eqnarray}
 & {L}_0P_{,0}+(\alpha a_0+\beta b_0+\gamma c_0)P_{,1}+(\alpha b_0+\beta d_0+\gamma e_0)P_{,2} &
\nonumber\\
&
+\mbox{}	(\alpha c_0+\beta e_0+\gamma f_0)P_{,3}+2{L}_0{}'=0,
&
 \label{eq303} 
\end{eqnarray}
here and in the sequel, a comma  means a partial differentiation, and a prime means differentiation with respect to a single variable on which the function depends. 

Using condition (\ref{norm_condition_3-0}) and finding the integrals of equation (\ref{eq303}), one can obtain expressions for the wave vector  and the energy density of radiation.
Below are listed all the solutions we have obtained for conformally Stackel space-times (3.0) type.

\subsubsection{CSS (3.0) type, case 1. (${L}_0\neq 0$).}
The wave vector of the radiation has the form:
\begin{eqnarray}
&
{L}_i=\left( L_0(x^0), \alpha, \beta, \gamma \right),
\quad
\alpha, \beta, \gamma  - const,
&\\[1ex]
&
 {L}_0=\sqrt{\alpha^2a_0+2\alpha\beta b_0+\beta^2d_0+2\alpha\gamma c_0+2\beta\gamma e_0+\gamma^2 f_0}.
 &\nonumber
\end{eqnarray}
For the energy density of radiation we obtain the expression:
\begin{eqnarray}
&
{\varepsilon}=F(X,Y,Z){\Delta\sqrt{- \det g^{ij}}}/{{L}_0},
&
\\[1ex]
&
X=x^1-\int{(\alpha a_0+\beta b_0+\gamma c_0)}/{{L}_0}\,dx^0,
&\nonumber\\
&
Y=x^2-\int{(\alpha b_0+\beta d_0+\gamma e_0)}/{{L}_0}\,dx^0,
&\nonumber\\
&
Z=x^3-\int{(\alpha c_0+\beta e_0+\gamma f_0)}/{{L}_0}\,dx^0,
& \nonumber
\end{eqnarray}
where $ F (X, Y, Z) $ is an arbitrary function of its variables.

\subsubsection{CSS (3.0) type, case 2. (${L}_0= 0$).}
For this degenerate case there is an additional condition for the functions of the metric (\ref{metric-3-0}):
\begin{eqnarray}
&
 \alpha^2a_0+2\alpha\beta b_0+2\alpha\gamma c_0+\beta^2d_0+2\beta\gamma e_0+\gamma^2 f_0=0,
 & \nonumber\\
&
\alpha, \beta, \gamma  - const.
&
\end{eqnarray}
The wave vector takes the form:
\begin{equation}  
{L}_i=(0,\alpha, \beta, \gamma).
\end{equation} 
For the energy density of radiation we obtain the expression:
\begin{eqnarray}
&
 	{\varepsilon}=F(x^0,Y,Z){\Delta\sqrt{- \det g^{ij}}}
&\\[1ex]
&
Y={x^1}/{(\alpha a_0+\beta b_0+\gamma c_0)}
-{x^2}/{(\alpha b_0+\beta d_0+\gamma e_0)},
 & \nonumber\\
&
Z={x^1}/{(\alpha a_0+\beta b_0+\gamma c_0)}
-{x^3}/{(\alpha c_0+\beta e_0+\gamma f_0)},
 & \nonumber
\end{eqnarray}
where $F(x^0,Y,Z)$  is  an arbitrary function of its variables.

\subsubsection{CSS (3.0) type, case 3. (${L}_0=0$, $\alpha a_0+\beta b_0+\gamma c_0=0$).}

In this degenerate case, when one of the terms in (\ref{eq303}) additionally vanishes together with $ {L}_0 $ (for example, the term with $ P_{, 1} $), we obtain additional conditions for the metric (\ref{metric-3-0}):
\begin{equation}  
\alpha a_0+\beta b_0+\gamma c_0=0,
\quad
\alpha, \beta, \gamma  - const,
 \end{equation} 
\begin{equation} 
 \alpha^2a_0+2\alpha\beta b_0+2\alpha\gamma c_0+\beta^2d_0+2\beta\gamma e_0+\gamma^2 f_0=0.
 \end{equation} 
The wave vector of the radiation has the form:
\begin{equation}  
{L}_i=(0,\alpha, \beta, \gamma).
\end{equation} 
For the energy density of radiation we obtain the expression:
\begin{equation} 
 	{\varepsilon}=F(x^0,x^1,Z){\Delta\sqrt{- \det g^{ij}}}
\end{equation} 
\[
Z=x^2 (\alpha c_0+\beta e_0+\gamma f_0) 
- x^3 (\alpha b_0+\beta d_0+\gamma e_0),
\]
where $F(x^0,x^1,Z)$ is an arbitrary function of its variables.

\subsection{Conformally Stackel space-times (3.1) type}
Conformally Stackel space-times (3.1) type admits 3 commuting Killing vectors $Y^i_{(p)}$ ($p=1,3$), but
 $rank\,|Y^i_{(p)}g_{ij}Y^j_{(q)}|=2$.
In a privileged coordinate system the metric of a conformally Stackel space-times (3.1) type can be written in the following form, where the variable $ x^0 $ is a null ("wave-like") variable:
\begin{equation}  
g^{ij}=\frac 1\Delta\pmatrix{0&1&a_0&b_0\cr 1&0&0&0\cr a_0&0&c_0&f_0\cr b_0&0&f_0&d_0},
\label{metric-3-1}
\end{equation} 
where  $\Delta=\Delta(x^0,x^1,x^2,x^3)$ and  $a_0, b_0, c_0, d_0, f_0$ are functions of a variable $ x^0$.

The wave vector of the radiation has the form:
\begin{eqnarray}
&
{L}_0={L}_0(x^0),\quad {L}_1=\alpha,\quad {L}_2=\beta,\quad {L}_3=\gamma,
&\\
&
\alpha, \beta, \gamma  - const.
\nonumber
\end{eqnarray}
The system of equations (\ref{law_of_conservation}), (\ref{eq_norm0}) takes the form:
\begin{equation}  
\beta^2 c_0+2\beta\gamma f_0+\gamma^2 d_0+2(\alpha+\beta a_0+\gamma b_0){L}_0=0,
\label{eq312}
\end{equation} 
\begin{eqnarray}
&
(\alpha+\beta a_0+\gamma b_0)P_{,0}+{L}_0P_{,1}+(a_0{L}_0+\beta c_0+\gamma f_0)P_{,2}
&\nonumber\\
& 
\mbox{}+(b_0{L}_0+\beta f_0+\gamma d_0)P_{,3}+2\beta a_0'+2\gamma b_0'=0.
&
\label{eq313}
\end{eqnarray}
From  equations (\ref{eq312})-(\ref{eq313}) we can obtain functional expressions for the wave vector and the energy density of radiation. Below all the solutions for conformally Stackel space-times (3.1) are listed.

\subsubsection{CSS (3.1) type, case 1. ($\alpha+\beta a_0+\gamma b_0\neq 0$).}
The wave vector of the radiation has the form:
\begin{equation}  {L}_i=\left( L_0(x^0), \alpha, \beta, \gamma \right),
\quad
\alpha, \beta, \gamma  - const,
\end{equation} 
 \begin{equation}  {L}_0=\frac{-(\beta^2 c_0+2\beta\gamma f_0+\gamma^2 d_0)}{2\,(\alpha+\beta a_0+\gamma b_0)}.
\end{equation} 
For the energy density of radiation we obtain the expression:
\begin{equation} 
	{\varepsilon}=F(X,Y,Z){\Delta \sqrt{- \det g^{ij}}}/{(\alpha+\beta a_0+\gamma b_0)},
\end{equation} 
\begin{eqnarray}
X&=&x^1-\int\frac{{L}_0}{(\alpha+\beta a_0+\gamma b_0)}\,dx^0,
\nonumber\\
Y&=&x^2-\int\frac{(a_0{L}_0+\beta c_0+\gamma f_0)}{(\alpha+\beta a_0+\gamma b_0)}\,dx^0,
\nonumber\\
Z&=&x^3-\int\frac{(b_0{L}_0+\beta f_0+\gamma d_0)}{(\alpha+\beta a_0+\gamma b_0)}\,dx^0,
\nonumber
\end{eqnarray}
where $F(X,Y,Z)$ is an arbitrary function of its variables. 

\subsubsection{CSS (3.1) type, case 2. ($L_0\neq 0$, $\alpha+\beta a_0+\gamma b_0= 0$). }
In this degenerate case  there are additional conditions for the metric (\ref{metric-3-1}):
\begin{equation} 
\alpha+\beta a_0+\gamma b_0= 0,
\quad
\alpha, \beta, \gamma  - const,
\end{equation} 
\begin{equation}  
\beta^2 c_0+2\beta\gamma f_0+\gamma^2 d_0=0.
\end{equation} 
The wave vector of the radiation has the form:
\begin{equation}  {L}_i=\left( L_0(x^0), \alpha, \beta, \gamma \right).
\end{equation} 
For the energy density of radiation we obtain the expression:
\begin{equation} 
	{\varepsilon}=F(x^0,Y,Z) \Delta \sqrt{- \det g^{ij}}.
\end{equation} 
\begin{eqnarray}
Y&=& x^1-\frac{x^2(a_0{L}_0+\beta c_0+\gamma f_0) }{L_0},
\nonumber\\
Z&=& x^1-\frac{x^3(b_0{L}_0+\beta f_0+\gamma d_0) }{L_0},
\nonumber
\end{eqnarray}
where $L_0(x^0)$ and $F(x^0,Y,Z)$  are arbitrary functions of their variables.

\subsubsection{CSS (3.1) type, case 3. ($L_0= 0$, $\alpha+\beta a_0+\gamma b_0= 0$).}
In this degenerate case  there are additional conditions for the metric  (\ref{metric-3-1}):
\begin{equation} 
\alpha+\beta a_0+\gamma b_0= 0,
\quad
\alpha, \beta, \gamma  - const,
\end{equation} 
\begin{equation}  
\beta^2 c_0+2\beta\gamma f_0+\gamma^2 d_0=0.
\end{equation} 
The wave vector of the radiation has the form:
\begin{equation}  
{L}_i=\left( 0, \alpha, \beta, \gamma \right).
\end{equation} 
For the energy density of radiation we obtain the expression:
\begin{equation} 
	{\varepsilon}=F(x^0,x^1,Z) \Delta \sqrt{- \det g^{ij}}.
\end{equation} 
\[ 
Z=x^2(\beta f_0+\gamma d_0)-x^3(\beta c_0+\gamma f_0),
\]
where $F(x^0,x^1,Z)$  is an arbitrary function of its variables.

\subsection{Conformally Stackel space-times (2.0) type}
Conformally Stackel space-times (2.0) type admit two commuting Killing vectors. In the privileged coordinate system, the metric  can be written in the following form:
\begin{equation}  
g^{ij}=\frac 1\Delta\pmatrix{1&0&0&0\cr 0&{{\epsilon}} &0&0\cr 0&0&A&B\cr 0&0&B&C},
\label{metric-2-0}
\end{equation} 
\begin{eqnarray*}
&
 \Delta=\Delta(x^0,x^1,x^2,x^3),  
\quad {\epsilon} =\pm 1,
\quad
A=a_0(x^0)+a_1(x^1),
&\\
&
B=b_0(x^0)+b_1(x^1), 
\qquad
C=c_0(x^0)+c_1(x^1) .
&
\end{eqnarray*}
The wave vector of the radiation ${L}_i$ has the form:
\begin{eqnarray}
&
{L}_0={L}_0(x^0),\qquad {L}_1={L}_1(x^1),
&\\
&
{L}_2=\alpha,\qquad {L}_3=\beta,  \qquad  \alpha, \beta  - const.
&\nonumber
\end{eqnarray}
From the norm condition  (\ref{eq_norm0}) one can obtain:
\begin{eqnarray}
&
{{L}_0}^2+\alpha^2 a_0+2\alpha\beta b_0+\beta^2 c_0-\gamma=0,
\label{eq201}
&\\
&
{\epsilon} {{L}_1}^2+\alpha^2 a_1+2\alpha\beta b_1+\beta^2 c_1+\gamma=0. 
&
\end{eqnarray}
From the conservation law (\ref{law_of_conservation}) one can obtain the equation for the energy density of radiation:
\begin{eqnarray}
&
{L}_0P_{,0}+{\epsilon} {L}_1P_{,1}+(\alpha A+\beta B)P_{,2}+(\alpha B+\beta C)P_{,3}
&\nonumber\\
&
\mbox{}+2{L}_0'+2{\epsilon}  {L}_1'=0.
&
\end{eqnarray}
Below are listed all the solutions for conformally Stackel space-times (2.0) type.

\subsubsection{CSS (2.0) type, case 1. (${L}_0{L}_1\neq 0$):}
The wave vector of the radiation has the form:
\begin{eqnarray}
&
{L}_i=\left( L_0(x^0), {L}_1(x^1), \alpha, \beta\right),  \qquad  \alpha, \beta,\gamma  - const,
&\nonumber\\[1ex]
&
{L}_0=\sqrt{\gamma-\alpha^2 a_0-2\alpha\beta b_0-\beta^2 c_0},
&\nonumber\\
&
{L}_1=\sqrt{{\epsilon} (-\gamma-\alpha^2 a_1-2\alpha\beta b_1-\beta^2 c_1)}.
&
\end{eqnarray}
For the energy density of radiation we obtain the expression:
\begin{equation} 
{\varepsilon}=F(X,Y,Z)\,{\Delta\sqrt{-\det g^{ij}}}/({{L}_0\,{L}_1}),
\end{equation} 
\begin{eqnarray*}
&
X=\int 1/{{L}_0}\, dx^0- \int {\epsilon}/{{L}_1}\, dx^1,
&\\
&
Y=x^2-\int{(\alpha a_0+\beta b_0)}/{{L}_0}\, dx^0 - {\epsilon} \int{(\alpha a_1+\beta b_1)}/{{L}_1}\, dx^1,
&\\
&
Z=x^3-\int{(\alpha b_0+\beta c_0)}/{{L}_0}\, dx^0 - {\epsilon} \int{(\alpha b_1+\beta c_1)}/{{L}_1}\, dx^1,
&
\end{eqnarray*}
where $F(X,Y,Z)$ is an arbitrary function of its variables.

\subsubsection{CSS (2.0) type, case 2. (${L}_0=0, {L}_1\neq 0$):}
Let us consider the degenerate case when $ {L}_0=0$ or ${L}_1 = 0 $ and fo rcertainty we take $ {L}_0 = 0 $ and $ {L}_1 \neq 0 $, then the wave vector will take the following form:
\begin{equation} 
{L}_i=\left( 0, {L}_1(x^1), \alpha, \beta\right),  \qquad  \alpha, \beta, \gamma  - const,
\end{equation} 
\begin{equation} 
		{L}_1=\sqrt{{\epsilon} (-\gamma-\alpha^2 a_1-2\alpha\beta b_1-\beta^2 c_1)}.
\end{equation} 
In this degenerate case  there are additional conditions for the metric (\ref{metric-2-0}):
\begin{equation} 
\alpha^2 a_0+2\alpha\beta b_0+\beta^2 c_0-\gamma=0.
\end{equation} 
For the energy density of radiation we obtain the expression:
\begin{equation} 
{\varepsilon}=F(x^0,Y,Z)\,{\Delta\sqrt{-\det g^{ij}}}/{{L}_1},
\end{equation} 
\[
Y=x^2-\int{(\alpha a_1+\beta b_1)}/{{L}_1}\, dx^1,
\]
\[
Z=x^3- {\epsilon} \int{(\alpha b_1+\beta c_1)}/{{L}_1}\, dx^1,
\]
where $F(x^0,Y,Z)$  is an arbitrary function of its variables.

\subsubsection{CSS (2.0) type, case 3. (${L}_0={L}_1= 0$):}
In this degenerate case, there are additional conditions for the metric (\ref{metric-2-0}):
\begin{equation} 
\alpha^2 A+2\alpha\beta B+\beta^2 C=0,
\qquad  \alpha, \beta  - const.
\end{equation} 
The wave vector of the radiation has the form:
\begin{equation} 
{L}_i=\left( 0, 0, \alpha, \beta\right).
\end{equation} 
The energy density of the radiation has the form:
\begin{equation} 
{\varepsilon}=F(x^0,x^1,Z)\,\Delta\sqrt{-\det g^{ij}},
\end{equation} 
\[
Z=x^2 (\alpha B+\beta C) -x^3(\alpha A+\beta B),
\]
where $F(x^0,x^1,Z)$  is an arbitrary function of its variables.

\subsection{Conformally Stackel space-times (2.1) type}
Conformally Stackel space-times (2.1) type admits two commuting Killing vectors $Y^i_{(p)}$ ($p=1,2$), but
 $rank\,|Y^i_{(p)}g_{ij}Y^j_{(q)}|=1$.
In a privileged coordinate system the metric of a conformally Stackel space-times (2.1) type can be written in the following form, where $ x^1 $ is a null ("wave-like") variable:
\begin{equation}  
g^{ij}=\frac 1\Delta\pmatrix{1&0&0&0\cr 0&0&f_1(x^1)&1\cr 0&f_1(x^1)&A&B\cr 0&1&B&C},
\label{metric-2-1}
\end{equation} 
\begin{eqnarray*}
&
\Delta=\Delta(x^0,x^1,x^2,x^3), \quad A=a_0(x^0)+a_1(x^1),
&\\
&
 B=b_0(x^0)+b_1(x^1),\quad C=c_0(x^0)+c_1(x^1).
 &
 \end{eqnarray*}
The wave vector of radiation $L_i$ has the form:
\begin{eqnarray}
&
{L}_0={L}_0(x^0),\quad {L}_1={L}_1(x^1),
&\nonumber\\
&
{L}_2=\alpha,\quad {L}_3=\beta,  \quad  \alpha, \beta, \gamma  - const.
&
\end{eqnarray}
From the norm condition (\ref{eq_norm0}) one can obtain:
\begin{eqnarray}
&
\gamma={{L}_0}^2+\alpha^2 a_0+2\alpha\beta b_0+\beta^2 c_0,
&\\
&
-\gamma=2(\alpha f_1+\beta){L}_1+\alpha^2 a_1+2\alpha\beta b_1+\beta^2 c_1.
\label{eq211}
&
\end{eqnarray}
From the conservation law (\ref{law_of_conservation}), one can obtain the equation for the energy density of radiation:
\begin{eqnarray}
&
 {L}_0P_{,0}+(\alpha f_1+\beta ) P_{,1}+(\alpha A+\beta B+f_1 {L}_1)P_{,2}
&\nonumber\\
&
\mbox{}+(\alpha B+\beta C+{L}_1)P_{,3}+2\alpha f_1'+2{{L}_0}'=0.
\label{eq212}
&
\end{eqnarray}
From the system of equations  (\ref{eq211})--(\ref{eq212}) we can obtain functional expressions through the metric  for the wave vector and the radiation energy density. 
Below all the solutions for conformally Stackel space-times (2.1) type are listed.

\subsubsection{CSS (2.1) type, case 1. ${L}_0(\alpha f_1+\beta)\neq 0$:}
The wave vector of radiation has the form:
\begin{eqnarray}
&
{L}_i=\left( L_0(x^0), {L}_1(x^1), \alpha, \beta\right),  \qquad  \alpha, \beta,\gamma  - const,
&\nonumber\\[1ex]
& 
 {L}_0=\sqrt{\gamma-\alpha^2 a_0-2\alpha\beta b_0-\beta^2 c_0},
&\\
&
{L}_1={(-\gamma-\alpha^2 a_1-2\alpha\beta b_1-\beta^2 c_1)}/\big({2\,(\alpha f_1+\beta)}\big).
&\nonumber
\end{eqnarray}
The energy density of the radiation has the form:
\begin{equation} 
{\varepsilon}=F(X,Y,Z)\,{\Delta\sqrt{-\det g^{ij}}}/{\big({L}_0(\alpha f_1+\beta)\big)},
\end{equation} 
\[
X=\int\frac{dx^0}{{L}_0}- \int\frac{dx^1}{(\alpha f_1+\beta)},\]
\[
Y=x^2-\int \frac{(\alpha a_0+\beta b_0)}{{L}_0}\,dx^0 -\int\frac{(\alpha a_1+\beta b_1+f_1L_1)}{\alpha f_1+\beta}\, dx^1,
\]
\[
Z=x^3-\int \frac{(\alpha b_0+\beta c_0)}{{L}_0}\,dx^0 -\int\frac{(\alpha b_1+\beta c_1+L_1)}{\alpha f_1+\beta}\, dx^1,
\]
where $F(X,Y,Z)$ is an arbitrary function of its variables.

\subsubsection{CSS (2.1) type, case 2. $(\alpha f_1+\beta)= 0$, $L_0\neq 0$.}
In this degenerate case, there are additional conditions for the metric (\ref{metric-2-1}):
\begin{equation} 
a_1=f_1=0.
\end{equation} 
The wave vector of radiation has the form ($\beta=\gamma=0$):
\[
{L}_i=\left( L_0(x^0), {L}_1(x^1), \alpha, 0\right), 
\]
\begin{equation} 
{L}_0=\alpha \sqrt{- a_0},
\quad
L_1=\sigma-\alpha b_1,
\quad
\alpha, \sigma - const.
\end{equation} 
The energy density of the radiation has the form:
\begin{equation}  
{\varepsilon}=F(x^1,Y,Z)\,{\Delta\sqrt{-\det g^{ij}}}/{L_0(x^0)},
\end{equation} 
\[
Y=\alpha x^2+\int{L_0}\,dx^0,
\quad
Z= x^3 -\int {(\sigma+\alpha b_0)}/{L_0}\,dx^0,
\]
where $F(x^1,Y,Z)$ is an arbitrary function of its variables.

\subsubsection{CSS (2.1) type, case 3. ${L}_0= 0$, $(\alpha f_1+\beta)\neq 0$.}
In this degenerate case, there are additional conditions for the metric  (\ref{metric-2-1}):
\begin{eqnarray}
&
\alpha^2 a_0+2\alpha\beta b_0+\beta^2 c_0=\gamma,
&\\
&
p(\alpha a_0+\beta b_0)+q (\alpha b_0+\beta c_0)=r,
&\\
&
\alpha, \beta,  \gamma, p, q, r \mbox{ -- const}.
&\nonumber
\end{eqnarray}
The wave vector of radiation has the form:
\begin{eqnarray}
&
{L}_i=\left( 0, {L}_1(x^1), \alpha, \beta\right), 
&\\
&
{L}_1= {(-\alpha^2 a_1-2\alpha\beta b_1-\beta^2 c_1-\gamma)}/{(2\,(\alpha f_1+\beta))}.
&\nonumber
\end{eqnarray}
The energy density of the radiation has the form:
\begin{equation}  {\varepsilon}=F(x^0,Y,Z)\,{\Delta\sqrt{-\det g^{ij}}}/{(\alpha f_1+\beta)},
\end{equation} 
\[
Y=\alpha x^2+\beta x^3 +\int L_1(x^1)\,dx^1,
\]
\[
Z=px^2+qx^3
\]
\[
\mbox{}-\int\frac{r+ p(\alpha a_1+\beta b_1+f_1L_1)+q(\alpha b_1+\beta c_1+L_1)}{(\alpha f_1+\beta)}\,dx^1,
\]
where $F(x^0,Y,Z)$  is an arbitrary function of its variables.

\subsubsection{CSS (2.1) type, case 4. ${L}_0= 0$, $L_1\neq 0$, $\alpha f_1+\beta= 0$.}
The wave vector of radiation has the form:
\begin{equation} 
{L}_i=\left( 0, {L}_1(x^1), 0, 0\right),
\qquad
 \alpha= \beta=\gamma=0.
\end{equation} 
The energy density of the radiation has the form:
\begin{equation}  {\varepsilon}=F(x^0,x^1,Z)\,\Delta\sqrt{-\det g^{ij}},\end{equation} 
\[
Z=x^2 -x^3f_1(x^1),
\]
where $L_1(x^1)$ and $F(x^0,x^1,Z)$  are arbitrary functions of their variables.


\subsection{Conformally Stackel space-times (1.0) type}
Conformally Stackel space-times (1.0) type admits one Killing vector, and
in a privileged coordinate system the metric of this space-time can be written in the following form:
\begin{equation}  
g^{ij}=\frac 1\Delta\pmatrix{\Omega &0&0&0\cr 0&V^1&0&0\cr 0&0&V^2&0\cr 0&0&0&V^3},
\label{metric-1-0}
\end{equation} 
\begin{eqnarray*}
&
 \Delta=\Delta(x^0,x^1,x^2,x^3), 
&\\
&
 V^1=t_2(x^2)-t_3(x^3),\qquad V^2=t_3(x^3)-t_1(x^1),
 &\\
&
 V^3=t_1(x^1)-t_2(x^2),
  \quad
 \Omega=\omega_\nu(x^\nu) V^\nu, 
 \quad 
\mu,\,\nu=1...3.
&
\end{eqnarray*}
The wave vector of radiation has the form:
\[ {L}_i=\Big(\alpha,{L}_1(x^1),{L}_2(x^2),{L}_3(x^3)\Big),\qquad \alpha=const.\]
From equations  (\ref{law_of_conservation}), (\ref{eq_norm0}) we have:
\begin{equation}  \Omega \alpha^2 + V^\mu {{L}_\mu}^2=0,\label{eq101}\end{equation} 
\begin{equation}   \alpha \Omega P_{,0}+ V^\mu({L}_\mu P_{,\mu}+2{L}_\mu')=0. \label{eq101-1}\end{equation} 
From this system of equations we obtain functional expressions  for the wave vector and the radiation energy density through the metric. Below are listed all the solutions for conformally Stackel space-times (1.0) type.

\subsubsection{CSS (1.0) type, case 1. ${L}_1{L}_2{L}_3\neq 0$.}

The wave vector of radiation has the form:
\begin{eqnarray}
&
{L}_i=\left(\alpha,{L}_1(x^1),{L}_2(x^2),{L}_3(x^3)\right),
\quad
\alpha,\beta,\gamma \mbox{ -- const}.
&\nonumber\\[1ex]
&
{L}_\mu=\sqrt{-\alpha^2\omega_\mu+\beta t_\mu+\gamma},
\quad
\mu, \nu=1...3.
&
\end{eqnarray}
The energy density of the radiation has the form:
\begin{equation}  
{\varepsilon}=F(X,Y,Z)\,{\Delta\sqrt{-\det g^{ij}}}/{({L}_1\,{L}_2\,{L}_3)},
\end{equation} 
\begin{eqnarray*}
&
 X=x^0 -\alpha\sum_\mu \int({\omega_\mu}/{{L}_\mu})\, dx^\mu,
&\\
&
 Y=\sum_\mu  \int ({t_\mu}/{{L}_\mu})\, dx^\mu ,
\qquad
 Z= \sum_\mu \int (1/{{L}_\mu}){\,dx^\mu} ,
 &
\end{eqnarray*}
 where $F(X,Y,Z)$  is an arbitrary function of its variables.

\subsubsection{CSS (1.0) type, case 2. ${L}_1=0, {L}_2{L}_3 \neq 0$.}
In this degenerate case, when one of the components of the wave vector $ {L}_\mu $ becomes zero (for definiteness, let $ L_1 = 0 $, $ {L}_2 {L}_3 \neq 0 $) we have an additional condition for the metric (\ref{metric-1-0}):
\begin{equation} 
\alpha^2\omega_1=\beta t_1+\gamma,
\qquad
\mbox{$\alpha,\beta,\gamma$ -- const}.
\end{equation} 
The wave vector of radiation has the form:
\begin{equation}  
{L}_i=\Big(\alpha,0,{L}_2(x^2),{L}_3(x^3)\Big),
\end{equation} 
\[
{L}_2=\sqrt{-\alpha^2\omega_2+\beta t_2+\gamma},
\quad
{L}_3=\sqrt{-\alpha^2\omega_3+\beta t_3+\gamma}.
\]
The energy density of the radiation has the form:
\begin{equation}  
{\varepsilon}=F(x^1,Y,Z)\,\Delta\sqrt{-\det g^{ij}}/\prod_{\mu\neq 1}{L}_\mu,
\end{equation} 
\[
 Y=\alpha x^0 -\alpha^2\sum_{\mu\neq 1}  \int ({\omega_\mu}/{{L}_\mu})\, dx^\mu +\beta\sum_{\mu\neq 1}  \int ({t_\mu}/{{L}_\mu})\, dx^\mu,
\]
\[
Z=\sum_{\mu\neq 1} \int(1/{{L}_\mu}){\,dx^\mu},
\]
where $F(x^1,Y,Z)$  is an arbitrary function of its variables.

\subsubsection{CSS (1.0) type, case 3. $L_1\neq 0, {L}_2=0, {L}_3= 0$.}
In this degenerate case, when two of the components of the wave vector $ {L}_\mu $ becomes zero (for definiteness, let $ L_1 \neq 0 $, $ {L}_2 ={L}_3= 0 $) we have additional conditions for the metric (\ref{metric-2-1}):
\begin{equation} 
\alpha^2\omega_2=\beta t_2+\gamma,
\quad
\alpha^2\omega_3=\beta t_3+\gamma,
\end{equation} 
where
 $\alpha$, $\beta$, $\gamma$ -- const.
 
The wave vector of radiation has the form:
\begin{equation}  
{L}_i=\left(\alpha,{L}_1(x^1),0,0\right),
\quad
{L}_1=\sqrt{\beta t_1-\alpha^2\omega_1+\gamma}.
\end{equation} 
The energy density of the radiation has the form:
\begin{equation}  
{\varepsilon}=F(x^2,x^3, Y)\,{\Delta\sqrt{-\det g^{ij}}}/{{L}_1(x^1)},
\end{equation} 
\[
Y=\alpha x^0+\int L_1(x^1) \,dx^1,
\]
where $F(x^2,x^3,Y)$  is an arbitrary function of its variables.
 
\subsection{Conformally Stackel space-times (1.1) type}
Conformally Stackel space-times (1.1) type admits one Killing vector.
In a privileged coordinate system the metric of a conformally Stackel space-times (1.1) type can be written in the following form, where $ x^1 $ is a null ("wave-like") variable:
\begin{equation}  
g^{ij}=\frac 1\Delta\pmatrix{\Omega&V^1&0&0\cr V^1&0&0&0\cr 0&0&V^2&0\cr 0&0&0&V^3},
\label{metric-1-1}
\end{equation} 
\begin{eqnarray*}
&
 \Delta=\Delta(x^0,x^1,x^2,x^3), 
&\\
&
 V^1=t_2(x^2)-t_3(x^3),\quad V^2=t_3(x^3)-t_1(x^1),
&\\
&
V^3=t_1(x^1)-t_2(x^2),
\quad 
\Omega=\omega_\mu(x^\mu) V^\mu,
\quad
\mu,\nu=1...3.
\end{eqnarray*}
The wave vector of radiation has the following "separated"  form:
\begin{equation}  {L}_i=\Big(\alpha,{L}_1(x^1),{L}_2(x^2),{L}_3(x^3)\Big),\quad \alpha - const.\end{equation} 
From equations  (\ref{law_of_conservation}), (\ref{eq_norm0}) we have:
\begin{equation}  \alpha V^1(2 {L}_1+\alpha\omega_1)+V^2({L}_2{}^2+\alpha^2\omega_2)+
	V^3({L}_3{}^2+\alpha^2\omega_3)=0,\label{eq111}\end{equation} 
\[ 
(V^1{L}_1+\alpha\Omega)P_{,0}+ \alpha V^1P_{,1}+ {L}_2V^2P_{,2}+ {L}_3V^3P_{,3}
\]
\begin{equation} 
\mbox{}+ 2V^2{L}_2'+ 2V^3{L}_3' = 0.\label{eq112}
\end{equation} 
From this system of equations we obtain functional expressions  for the wave vector and the radiation energy density through the metric. Below are listed all the solutions for conformally Stackel space-times (1.1) type.

\subsubsection{CSS (1.1) type, case 1. $\alpha L_2 L_3\neq 0$.}

The wave vector of radiation has the form:
\[  
{L}_0=\alpha,
\quad
{L}_1=\frac 1{2\alpha}(\beta t_1 -\alpha^2\omega_1+\gamma),
\quad
\alpha,\beta, \gamma - const,
\]
\begin{equation} 
{L}_2=\sqrt{\beta t_2-\alpha^2\omega_2+\gamma},
\quad
{L}_3=\sqrt{\beta t_3-\alpha^2\omega_3+\gamma}.\end{equation} 
The energy density of the radiation has the form:
\begin{equation}  
{\varepsilon}=F(X,Y,Z)\,{\Delta\sqrt{-\det g^{ij}}}/{({L}_2\,{L}_3)},
\end{equation} 
\[
 X=x^0 -\frac{1}{\alpha} \int({L}_1+\alpha \omega_1)\,dx^1  -\alpha \left( \int\frac{\omega_2}{{L}_2}\,dx^2 +\int\frac{\omega_3}{{L}_3}\,dx^3 \right),
\]
\[
 Y=-\frac{1}{\alpha}\int t_1\, dx^1  +\int\frac{t_2}{{L}_2}\,dx^2+\int\frac{t_3}{{L}_3}\,dx^3 ,
\]
\[
 Z= \frac{x^1}{\alpha}+\int\frac{dx^2}{{L}_2}+\int\frac{dx^3}{{L}_3},
\]
where $F(X,Y,Z)$  is an arbitrary function of its variables.

\subsubsection{CSS (1.1) type, case 2. $\alpha= 0$.}
In this degenerate case, when $ {L}_0 = 0 $, we obtain an additional condition for the metric (\ref{metric-1-1}):
\begin{equation} 
t_1=0.
\end{equation} 
The wave vector of the radiation $L_i$ has the following form ($\gamma=0$):
\begin{equation} 
{L}_0=0,
\quad
{L}_1=L_1(x^1),
\quad
{L}_2=\sqrt{\beta t_2},
\quad
{L}_3=\sqrt{\beta t_3}.
\end{equation} 
The energy density of the radiation has the form:
\begin{equation} 
{\varepsilon}=F(x^1,Y,Z)\,{\Delta\sqrt{-\det g^{ij}}}/{({L}_2\,{L}_3)},
\end{equation} 
\[
 Y=\int\frac{t_2}{{L}_2}\,dx^2+\int\frac{t_3}{{L}_3}\,dx^3 ,
\quad
 Z=\frac{x^0}{L_1}+ \int\frac{dx^2}{{L}_2}+\int\frac{dx^3}{{L}_3},
\]
where $F(x^1,Y,Z)$ and $L_1(x^1)$ are arbitrary functions of its variables.

\subsubsection{CSS (1.1) type, case 3. $\alpha L_2\neq 0, L_3=0$.}

In the degenerate case, when one of the components of the wave vector $L_3$ becomes zero (similarly for $ L_2 $ with the replacement of the indices 3 by 2) we have  additional conditions for the metric (\ref{metric-2-1}):
\begin{equation} 
\alpha^2\omega_3=\beta t_3+\gamma, \qquad \alpha, \beta, \gamma - const,
\end{equation} 
\begin{equation} 
(p\alpha-q\beta)\,t_3=0,
\qquad
p,q - const.
\end{equation} 
The wave vector of the radiation $L_i$ has the following form:
\begin{equation} 
{L}_0=\alpha,
\qquad
{L}_1=\frac 1{2\alpha}(\beta t_1 -\alpha^2\omega_1+\gamma),
\end{equation} 
\[
{L}_2=\sqrt{\beta t_2-\alpha^2\omega_2+\gamma},
\qquad
{L}_3=0.
\]
The energy density of the radiation has the form:
\begin{equation} 
{\varepsilon}=F(x^1,Y,Z)\,{\Delta\sqrt{-\det g^{ij}}}/{{L}_2},
\end{equation} 
\[
Y=\alpha x^0+\int L_1\,dx^1+\int L_2\,dx^2,
\]
\[
Z=qx^0+\frac{1}{\alpha}\int \Big( q(\gamma/\alpha  -\alpha\omega_1 -L_1)+pt_1\Big)dx^1
\]
\[\mbox{}+
\int\frac{q(\gamma/\alpha-\alpha\omega_2)+pt_2}{L_2}\,dx^2,
\]
 where $F(x^1,Y,Z)$  is an arbitrary function of their variables.


\subsection{Conformally Stackel space-times  (0.0) type}
The metric of conformally Stackel space-times (0.0) type 
in a privileged coordinate system 
can be written in the following form:
\begin{equation} 
g^{ij}=\frac 1\Delta\pmatrix{V^0 &0&0&0\cr 0&V^1&0&0\cr 0&0&V^2&0\cr 0&0&0&V^3},
\label{metric-0-0}
\end{equation} 
\begin{eqnarray*}
&
\Delta=\Delta(x^0,x^1,x^2,x^3), 
&\\
&
V^0=a_1(b_2-b_3)+a_2(-b_1+b_3)+a_3(b_1-b_2),
&\\
&
V^1= a_0(-b_2+b_3)+a_2(b_0-b_3)+a_3(-b_0+b_2),
&\\
&
V^2=a_0(b_1-b_3)+a_1(-b_0+b_3)+a_3(b_0-b_1),
&\\
&
 V^3=a_0(-b_1+b_2)+a_1(b_0-b_2)+a_2(-b_0+b_1),
 &
\end{eqnarray*}
where functions $ a $, $ b $, $ c $ are functions of only one variable whose index corresponds to the lower index of the function. For example $ a_0 = a_0 (x ^ 0) $, $ b_1 = b_1 (x ^ 1) $, etc.

The wave vector of the radiation in a privileged coordinate system  has the following "separated" \, form:
\[ {L}_i=\Big({L}_0(x^0),{L}_1(x^1),{L}_2(x^2),{L}_3(x^3)\Big).\]
From the law of energy-momentum conservation  (\ref{law_of_conservation})  and condition (\ref{eq_norm0}) we have:
\begin{equation}  
V^i{L}_i{}^2=0,
\qquad
i=0...3,
\end{equation} 
\begin{equation}  V^i({L}_iP_{,i}+2{L}_i')=0.
\end{equation} 
From this system of equations we obtain functional expressions  for the wave vector and the radiation energy density through the metric. Below are listed all the solutions for conformally Stackel space-times (0.0) type.

\subsubsection{CSS (0.0) type, case 1. ${L}_0{L}_1{L}_2{L}_3\neq 0$.}
The wave vector of the radiation $L_i$ has the following form ($\alpha$, $\beta$, $\gamma$ -- const):
\begin{equation} 
{L}_i=\sqrt{\alpha a_i+\beta b_i+\gamma},
\qquad
i,j \mbox{ = 0...3.}
\end{equation} 
The energy density of the radiation has the form:
\begin{equation}  
{\varepsilon}=F(X,Y,Z)\,{\Delta\sqrt{-\det g^{ij}}}/{({L}_0\,{L}_1\,{L}_2\,{L}_3)},
\end{equation} 
\[
X=\sum_i \int\frac{dx^i}{{L}_i},
\quad
Y=\sum_i \int\frac{a_i}{{L}_i}\,dx^i,
\quad
Z=\sum_i \int\frac{b_i}{{L}_i}\,dx^i,
\]
where $F(X,Y,Z)$  is an arbitrary function of its variables.

\subsubsection{CSS (0.0) type, case 2. ${L}_0=0, \quad {L}_1{L}_2{L}_3\neq 0$.}
In the degenerate case, when one of the components of the wave vector turns to zero (for definiteness a component with the number $ i=0 $, that is, $ {L}_0 = 0 $), then we obtain an additional condition for the metric (\ref{metric-0-0}):
\begin{equation} 
\alpha a_0+\beta b_0+\gamma=0,
\qquad
\mbox{$\alpha$, $\beta$, $\gamma$ -- const.}
\end{equation} 
For the wave vector of radiation ${L}_i$ we obtain:
\begin{equation} 
{L}_{0}=0,
\qquad
{L}_i=\sqrt{\alpha a_i+\beta b_i+\gamma},
\qquad
i\neq 0.
\end{equation} 
The energy density of the radiation has the form:
\begin{equation}  
{\varepsilon}=F(x^0,X,Y)\,\Delta\sqrt{-\det g^{ij}}/{\prod_{i\neq 0}{L}_i},
\end{equation} 
\[
X=\sum_{i\neq 0} \int\frac{dx^i}{{L}_i},
\qquad
Y=\sum_{i\neq 0} \int{L}_i\,dx^i,
\]
 where $F(x^0,X,Y)$  is an arbitrary function of its variables.

\subsubsection{CSS (0.0) type, case 3. ${L}_0={L}_1= 0, \quad {L}_2{L}_3\neq 0$.}
In the degenerate case, when the two components of the wave vector vanish (let it be the components $ {L}_0 =0$, $ {L}_1 = 0 $), then we obtain additional conditions for the metric 
 ($\alpha$, $\beta$, $\gamma$ -- const):
\begin{equation} 
\alpha a_0+\beta b_0+\gamma=0,
\quad
\alpha a_1+\beta b_1+\gamma=0.
\end{equation} 
For the wave vector of radiation  we have:
\[
{L}_i=\Big(0,0,L_2(x^2), L_3(x^3)\Big),
\]
\begin{equation} 
{L}_2=\sqrt{\alpha a_2+\beta b_2+\gamma},
\quad
{L}_3=\sqrt{\alpha a_3+\beta b_3+\gamma}.
\end{equation} 
The energy density of the radiation has the form:
\begin{equation}  
{\varepsilon}=F(x^0,x^1,Y)\,\frac{\Delta\sqrt{-\det g^{ij}}}{{L}_2{L}_3},
\end{equation} 
\[
Y= \int{L}_2\,dx^2 + \int{L}_3\,dx^3,
\]
 where $F(x^0,x^1,Y)$  is an arbitrary function of its variables.

Note that  for Stackel space-times (0.0) type the three components of the wave vector of the radiation can not be zero, since this leads to a violation of the norm condition.

\section{Conclusion}

In the paper we obtain and enumerate all solutions of the equations of the energy-momentum conservation law of pure radiation for models of space-times that admit separation of variables in the eikonal equation.

The forms of the energy-momentum tensor of pure radiation (energy density and wave vector of radiation) for all types of space-times under consideration are obtained. In privileged coordinate systems (where the separation of variables is admitted), the energy density and the wave vector of the radiation are determined through functions of the space-time metric.

The results obtained can be used to construct integrable models for various metric theories of gravitation, including for comparing similar models in the theory of gravity of Einstein and in modified theories of gravity.

\begin{acknowledgments}
Research partially supported by the Ministry of Education and Science of Russia under contract 3.1386.2017/4.6.
\end{acknowledgments}

\end{document}